
\baselineskip=2\baselineskip
\magnification 1200
\rightline{RU-94-33}
\vskip 0.4in
\centerline{\bf Boundary S-matrix for the Integrable}
\centerline{\bf q-Potts Model}
\vskip 0.6in
\centerline{Leung Chim \footnote{$^\dagger$}{E-mail:
CHIM@ruhets.rutgers.edu}}
\vskip 0.1in
\centerline{Department of Physics and Astronomy}
\centerline{Rutgers University}
\centerline{P.O.Box 849, Piscataway, NJ 08855-0849}
\vskip 0.4in
\centerline{\bf Abstract}

The 2D off-critical q-state Potts model with boundaries was studied as a
factorizable relativistic scattering theory. The scattering S-matrices
for particles reflecting off the boundaries were obtained for the
cases of ``fixed'' and ``free'' boundary conditions. In the Ising limit,
the computed results agreed with recent work[5].
\vfill\eject

{\bf Introduction}

In the past decade, many advances were made in 2 dimensional statistical
mechanic systems by applying the idea of conformal invariance[1]. Their
critical points are well described by conformal field theory (CFT) and
their universality can be classified by a Virasoro central charge $c$[2].
When such a model is perturbed off their critical point, the conformal
symmetry is broken and the theory develops finite correlation lengths.
However, for certain perturbations, residue symmetries survives in the form
of an infinite set of commuting integrals of motion and renders the theory
integrable[3].
Such ``perturbed conformal field theories'' can often be described by a
relativistic scattering theory of massive particles where the S-matrix is
factorizable. All physical information about the field theory can then be
obtained from the S-matrix by constructing correlation functions using the
form-factors method[4].

Recently, statistical systems with boundaries were studied using
the above method[5]. It was found that one can choose certain
boundary conditions which perserves the integrability of the bulk theory. Such
integrable boundary conditions can be represented by the boundary S-matrix
describing the scattering of particles from the boundary. In this work, such
boundary S-matrices are obtained for the q-state Potts model (for $0<q<4$) with
free and fixed boundary conditions.

{\bf Integrable q-states Potts Model}

In the lattice q-Potts model[6], the spins $a(x)$ at the sites of the lattice
are
allowed to be in one of the $q$ possible states $(1,2,...,q)$. The partition
function has the form
$$Z = \sum_{\lbrace a(x) \rbrace}
\prod_{(x,y)=<nn>}(1+K\delta_{a(x),a(y)}), \eqno(1)$$
which is invariant under the permutation group $S_q$. The phase transition
point
of this model occurs at
$$K = K_c = \sqrt{q}. \eqno(2)$$
For $0<q<4$, this is a second order phase transition, and its critical point is
described by the CFT with central charge[7]
$$c = 1 - {6\over{p(p+1)}} \quad where \quad \sqrt q =
2sin({\pi \over 2}{{p-1}\over{p+1}}). \eqno(3)$$
The energy density $\epsilon (x)$ of the Potts model then corresponds to the
degenerate primary field $\Phi_{(2,1)}$ with dimension
$$\qquad \Delta_\epsilon =
{1\over 4} + {3\over 4p}. \eqno(4)$$
This field is also a relevent operator in the theory, and preturbation of the
CFT action by this field leads to an off-critical theory with action
$$A_{q,\tau}=A_{CFT(c)}+\tau \int \epsilon(x){d^2}x \eqno(5)$$
where
$$\tau = {{K_c - K}\over K_c}, \eqno(6)$$
and $A_{CFT(c)}$ is the ``action'' of the CFT with central charge $c$. This
action
describes the scaling domain of the q-Potts model, and is shown in [8] to be an
integrable field theory (ie possess nontrival higher-spin local integrals of
motion). Following [8], we consider the low temperature phase $K > K_c$ of this
theory. The field theory (5) has $q$ degenerate vacua $\mid a\rangle$; $a =
1,...,q$,
with $S_q$ acting by permutations of these vacuum states. Its particle content
must
then contain $q(q-1)$ ``kinks'' $A_{ab}; \quad a,b=1,...,q; \quad a \neq b$,
which
corresponds to the domain wall separating vacua $a$ and $b$. The antiparticles
can
be identified by $\bar A_{ab} = A_{ba}$. The masses of these particles are
taken
to be equal to $M \sim {\mid \tau \mid}^{1\over{1-\Delta_{\epsilon}}}$.

The scattering of these asymptotic particles is governed by the S-matrix and
the
integrability of (5) implies that this S-matrix is factorizable. Recall that
the
energy-momentum of particles can be parametrized by their rapidity $\theta$,
where
$$E=Mch\theta,\quad P= Msh\theta. \eqno(7)$$
The asymptotic particles
states are generated by the ``particle creation operator'' $A_{ab}(\theta)$
satisfying
the quadratic commutation relations
$$A_{ab}(\theta_{1})A_{bc}(\theta_{2})=\sum_{d\neq a, d\neq
c}S_{ac}^{bd}(\theta_{12})
A_{ad}(\theta_{2})A_{dc}(\theta_{1}),\eqno(8)$$
where $\theta_{12}=\theta_{1}-\theta_{2}.$ The $S_q$ symmetric two-kink
S-matrix
elements (Fig.1) were computed in [8] to be (for $a \neq b\neq c\neq d$)
$$S_{ac}^{bd}(\theta)=S_{0}(\theta)={{sh(\lambda\theta)sh(\lambda(i\pi-\theta))}\over
{sh(\lambda(\theta-{{2\pi i}\over 3}))sh(\lambda({{i\pi}\over
3}-\theta))}}\Pi({{\lambda\theta}\over{i\pi}}); \eqno(9a)$$
$$S_{ac}^{bb}(\theta)=S_{1}(\theta)={{sin(\lambda{{2\pi}\over
3})}\over{sin(\lambda {\pi
\over 3})}}{{sh(\lambda(i\pi - \theta))}\over{sh(\lambda({{2\pi i}\over
3}-\theta))}}\Pi({{\lambda\theta}\over{i\pi}}); \eqno(9b)$$
$$S_{aa}^{bd}(\theta)=S_{2}(\theta)={{sin(\lambda{{2\pi}\over 3})}\over
{sin(\lambda {\pi
\over 3})}}{{sh(\lambda \theta)}\over {sh(\lambda(\theta-{{i\pi}\over
3}))}}\Pi({{\lambda\theta}\over{i\pi}}); \eqno(9c)$$
$$S_{aa}^{bb}(\theta)=S_{3}(\theta)={{sin(\lambda \pi)}\over {sin(\lambda{\pi
\over
3})}}\Pi({{\lambda\theta}\over{i\pi}}), \eqno(9d)$$
where
$$\lambda = {3\over 2}{{p-1}\over {p+1}} \eqno(10)$$
and
$$\Pi(x)=-
{{\Gamma(1-x)\Gamma(1-\lambda+x)
\Gamma({7\over 3}\lambda-x)\Gamma({4\over3}\lambda+x)}\over
{\Gamma(1+x)
\Gamma(1+\lambda-x)\Gamma({1\over3}\lambda+x)\Gamma({4\over 3}\lambda-x)}}
\prod_{k=1}^{\infty}{\Pi_{k}(x)\Pi_{k}(\lambda-x)}; $$
$$\Pi_{k}(x)=
{{\Gamma(1+2k\lambda-x)\Gamma(2k\lambda-x)\Gamma(1+(2k-{1\over
3})\lambda-x)\Gamma((2k+{7\over 3})\lambda-x)}\over{\Gamma(1+(2k+1)\lambda-x)
\Gamma((2k+1)\lambda-x)\Gamma(1+(2k-
{4\over 3})\lambda-x)\Gamma((2k+{4\over 3})\lambda-x)}}. \eqno(11)$$
The amplitudes $S_{1}(\theta)$ possess a ``bound-state'' pole at
$\theta = {2\pi i \over 3}$ (Fig.2a)
while $S_{2}(\theta)$ have the ``cross-channel'' pole at
$\theta = {\pi i \over 3}$ (Fig.2b).
Likewise $S_{0}(\theta)$ have both the above poles (Fig.2c) with residues
$$Res_{\theta ={{2\pi i}\over 3}}S_{0}(\theta)=
Res_{\theta ={{2\pi i}\over 3}}S_{1}(\theta)=$$
$$-Res_{\theta ={{i\pi}\over 3}}S_{0}(\theta)=
-Res_{\theta ={{i\pi}\over 3}}S_{2}(\theta)= if^2(\lambda),\eqno(12)$$
where
$$f(\lambda)=\sqrt{{1\over \lambda}sin({{2\pi}\over 3}\lambda)}
exp\lbrace \int_{0}^{\infty}{{(1-e^{-(1-{4\over
3}\lambda)t})(1-e^{-{2\over 3}\lambda t})(1-e^{{1\over 3}\lambda
t})e^{-\lambda t}}\over {2(1-e^{-t})(1+e^{-\lambda t})}}{dt\over
t}\rbrace \eqno(13)$$
is the ``three-kink coupling''(Fig.3).

As is shown in [8], the complete particle spectrum of field theory (5) for
$3<q<4$ is
quite complicated with the appearance of exotic excitation and bound states.
For
simplicity, we will restrict our attention to the range $0<q \le 3.$

{\bf Boundary S-Matrix}

It is natural to consider the Potts model in the presence of boundaries with
some
boundary conditions imposed on the boundary spins. In [5], it was shown that
certain
boundary conditions preserves the integrability of the bulk theory (ie an
infinite
subset of the bulk integrals of motion survives with the introduction of the
boundaries). For a relativistic scattering theory of massive particles, one can
associate these integrable boundary conditions with certain boundary S-matrix
which
describe the scattering of particles with the boundary.

To be more precise, consider the model defined on an semi-infinite plane
(say $x \in (-\infty,0], y \in (-\infty,\infty)$, the y-axis being the
booundary).
Let us suppose that there exists integrable boundary conditions for the Potts
model with the modified action
$$A = A_{q,\tau + CBC} + \int_{-\infty}^{\infty}{dy}{\Phi_{B}(y)}, \eqno(14)$$
where $A_{q,\tau + CBC}$ is the action (5) with certain conformal boundary
conditions (CBC), and $\Phi_{B}(y)$ is some relevant boundary operator[9,5].
One can think of (14) as a perturbation of CBC, and the corresponding Fock
states can be classified as asymptotic scattering states. In particular, the
boundary with boundary spins in the state ``a'' can be associated with a
stationary impenetrable particle $B_a$ of infinite mass at $x=0$. Then the
asymptotic n-kink scattering state can be written as the product
$$A_{a_{1}a_{2}}(\theta_{1})A_{a_{2}a_{3}}(\theta_{2})...
A_{a_{n-1}a_{n}}(\theta_{n-1})A_{a_{n}a}(\theta_{n})B_a, \eqno(15)$$
where the vacua $a_{1},a_{2},...,a_{n}$ satisfy the restrictions
$a_{i}\neq a_{i+1}$, and $a_{n}\neq a$.

If the initial ``in-state'' of scattering is the asymptotic state (15) with
$\theta_{1}>\theta_{2}>...>\theta_{n}>0$ (ie n kinks moving towards
the boundary of state ``a''), then in the infinite future, this state becomes
a superposition of the final ``out-states''. Integrability of (14) constraints
``out-states'' to have the form
$$A_{b_{1}b_{2}}(-\theta_1)...
A_{b_{n-1}b_{n}}(-\theta_{n-1})A_{b_{n}b}(-\theta_n)B_{b}, \eqno(16)$$
with $b_{i}\neq b_{i+1}$ and $b_{n}\neq b$. Thus we have the relation
$$A_{a_{1}a_{2}}(\theta_{1})A_{a_{2}a_{3}}(\theta_{2})...
A_{a_{n-1}a_{n}}(\theta_{n-1})A_{a_{n}a}(\theta_{n})B_a =$$
$$\sum_{b_1}...\sum_{b_n}\sum_{b}R_{a_{1}...a_{n}a}^
{b_{1}...b_{n}b}(\theta_{1},...,\theta_{n})
A_{b_{1}b_{2}}(-\theta_1)...
A_{b_{n-1}b_{n}}(-\theta_{n-1})A_{b_{n}b}(-\theta_n)B_{b}, \eqno(17)$$
which defines the n-kink  Boundary S-matrix. When n=1, we have the
simple commutation relation
$$A_{ba}(\theta)B_a = \sum_{c}R_{ba}^{c}(\theta)A_{bc}(-\theta)B_{c},
 \eqno(18)$$
where $R_{ba}^{c}(\theta)$ are elements of the boundary S-matrix for the
reflection of one particle off the boundary(Fig.4). It follows from the
factorizability
of the scattering that $R_{a_{1}...a_{n}a}^{b_{1}...b_{n}b}$
can be written as a product of bulk amplitudes $S_{ac}^{bd}$, and boundary
amplitudes $R_{ba}^{c}$. For example, when 2 kinks scatter off the
boundary, the amplitude for this scattering can be factorized in
two equivalent ways(Fig.5), leading to
$$\sum_{g}\sum_{f}R_{ba}^{f}(\theta_1)S_{cf}^{bg}(\theta_{1}+\theta_{2})
R_{gf}^{e}(\theta_2)S_{ce}^{gd}(\theta_{2}-\theta_{1})$$
$$=\sum_{g'}\sum_{f'}S_{ca}^{bg'}(\theta_{2}-\theta_{1})
R_{g'a}^{f'}(\theta_2)S_{cf'}^{g'd}(\theta_1+\theta_2)
R_{df'}^{e}(\theta_1), \eqno(19)$$
which is known as the ``boundary Yang-Baxter equation''[10].

As is known, the amplitudes $R_{ba}^{c}(\theta)$ have to satisfy
several conditions in addition to (19). Firstly we have the
``boundary unitarity condition''(Fig.6)
$$\sum_{c\neq b}R_{ba}^{c}(\theta)R_{bc}^{d}(-\theta) = \delta_{a}^{ d},
\eqno(20)$$
which is a direct analog to the unitarity condition for the bulk
S-matrix. To obtain the crossing symmetry condition for the boundary
scattering, it is necessary to use the ``cross amplitude''[5]
$$K^{abc}(\theta) = R_{ba}^{c}({{i\pi}\over 2}-\theta). \eqno(21)$$
As shown in [5], this amplitude $K^{abc}(\theta)$ has to satisfy the
so-called ``boundary cross-unitarity condition''
$$K^{abc}(\theta) = \sum_{d\neq a,c}S_{ca}^{db}(2\theta)K^{adc}(-\theta),
\eqno(22)$$
which is illustrated in Fig.7. Finally we have the ``boundary bootstrap
equation''[11] which describes the scattering of ``bound-state'' particles
with the boundary. In the bulk theory, the kink $A_{ab}$ can appear
as a bound-state particle in the two particle scattering (see Fig.2)
$$A_{ac}(\theta+{{i\pi}\over 3})A_{cb}(\theta-{{i\pi}\over 3})
 \rightarrow fA_{ab}(\theta) \eqno(23)$$
where $f$ is the 3-particle coupling in (13). Applying the algebras (8),
(18) and (23) to the ``in-state'' $A_{ac}(\theta+{{i\pi}\over 3})
A_{cb}(\theta-{{i\pi}\over 3})B_{b}$, we obtained the bootstrap
equation(Fig.8)
$$R_{ab}^{d}(\theta)=\sum_{f\neq c,e}\sum_{e\neq a,d}
R_{cb}^{f}(\theta-{{i\pi}\over 3})S_{af}^{ce}(2\theta)
R_{ef}^{d}(\theta+{{i\pi}\over 3}) \eqno(24)$$
for the $S_q$ symmetric Potts model.

Equations (19) through (24) allows the boundary S-matrix elements
$R_{ab}^{c}(\theta)$ to be determined up to some CDD factors[5]. For
boundary conditions which respect the $S_q$ symmetry, we can
expect $R_{ab}^{c}(\theta)$ to have a pole at $\theta={{i\pi}\over 6}$
(Fig.9a) with residue
$$R_{ab}^{c}(\theta) \simeq {i\over2}{{fg_{b}^{c}}\over
{\theta-{{i\pi}\over6}}}, \eqno(25)$$
where $g_{b}^{c}$ is the amplitude for coupling of the particle $A_{cb}$
to the boundary(Fig.10). Furthermore, if $g_{a}^{c}$, $g_{b}^{a}\neq 0$, the
element $R_{ab}^{c}(\theta)$ has another pole at $\theta={{i\pi}\over 2}$
(Fig.9b) where
$$R_{ab}^{c}(\theta) \simeq {i\over2}{{g_{a}^{c}g_{b}^{a}}\over
{\theta-{{i\pi}\over2}}}; \eqno(26)$$
this pole is shown in Fig.10. Of course the presence of the above poles
depends on the boundary condition, as we shall see when we consider
the two simplest cases: ``free'' and ``fixed'' boundary conditions. Both
cases are conformal boundary conditions (i.e. $\Phi_{B}(y)=0$)
and we conjecture that they
also preserves integrability in the off-critical Potts model.

{\bf Fixed Boundary Condition}

In this simple boundary condition, the boundary spins are all fixed
to one state, say ``a''. The corresponding boundary S-matrix element
$$R_{ba}^{a}(\theta) = R_{f}(\theta) \eqno(27)$$
satisfies the boundary Yang-Baxter equation (19) automatically. To
determine this amplitude, one appeals to the unitarity condition (20)
$$R_{f}(\theta)R_{f}(-\theta) = 1; \eqno(28)$$
and the crossing symmetry condition (22)
$$K_{f}(\theta)=[(q-2)S_{2}(2\theta)+S_{3}(2\theta)]
K_{f}(-\theta), \eqno(29)$$
where
$$K_{f}(\theta) = R_{f}({{i\pi}\over2}-\theta) \eqno(30)$$
is the crossing amplitude. Since all boundary states are fixed, we
do not expect $R_{f}(\theta)$ to possess any poles in the physical
domain $0\le \theta \le {{i\pi}\over 2}$. The solution to (28)
and (29) can be factorized as
$$R_{f}(\theta) = F_{0}(\theta)F_{1}({{\lambda \theta}\over {i\pi}}),
\eqno(31)$$
where $F_{1}(X)$ solves
$$F_{1}(X)=\Pi(\lambda-2X){{sin(2\pi(\lambda-X))}\over
{sin(2\pi({\lambda\over3}-X))}}F_{1}(\lambda-X); \eqno(32)$$
$$F_{1}(-X)F_{1}(X) = 1, \eqno(33)$$
and its minimal solution can be written as
$$F_{1}(X) = \prod_{k=1}^{\infty}{\Sigma_k(X)\over\Sigma_k(-X)}, \eqno(34a)$$
with
$$\Sigma_k(X) =$$
$${\Gamma[(4k-1)\lambda+2X]\Gamma[(4k-3)\lambda
+2X+1]\Gamma[(4k-3)\lambda+{\lambda\over3}+2X]
\Gamma[(4k-2)\lambda+{\lambda\over3}-2X]}\over
{\Gamma[4k\lambda+2X]\Gamma[4(k-1)\lambda+2X+1]
\Gamma[(4k-2)\lambda-{\lambda\over3}+2X+1]
\Gamma[(4k-1)\lambda-{\lambda\over3}-2X+1]}, \eqno(34b)$$
up to some CDD factors.

The factor $F_{0}(\theta)$ can be obtained from the fixed boundary bootstrap
equation (24)
$$R_{f}(\theta)=[S_{1}(2\theta)+(q-3)S_{0}(2\theta)]
R_{f}(\theta-{{i\pi}\over3})R_{f}(\theta+{{i\pi}\over3}), \eqno(35)$$
which reduces to
$$F_{0}(\theta)=-tan({\pi\over4}+{{i\theta}\over2})cot({\pi\over12}+{{i\theta}\over2})
cot({{5\pi}\over12}+{{i\theta}\over2})F_{0}(\theta-{{i\pi}\over3})
F_{0}(\theta+{{i\pi}\over3}), \eqno(36)$$
with simple solution
$$F_{0}(\theta) = -tan({\pi\over4}+{{i\theta}\over2}). \eqno(37)$$

In the Ising limit $(q=2, \lambda={3\over4})$, the boundary S-matrix
have the form
$$R_{f}(\theta) = itan({{i\pi}\over4}-{{\theta}\over2}), \eqno(38)$$
as obtained in [5] by field theoretic method.

In the other interesting limit $q=3$ ($\lambda=1$), suppose the boundary
are fixed in the state ``A'', the solution simplifies as
$$R_{f}(\theta) = -{sin({\pi\over3}+{{i\theta}\over2})\over
sin({\pi\over3}-{{i\theta}\over2})}. \eqno(39)$$

{\bf Free Boundary Condition}

In contrast to the fixed boundary condition, we have the ``free'' case
where the boundary spins can be in any one of the $q$ states. The
corresponding boundary S-matrix has to respect the $S_{q}$ symmetry
and the algebra (18) simplifies to
$$A_{ba}(\theta)B_{a}=R_{1}(\theta)A_{ba}(-\theta)B_{a}+\sum_{c\neq a,b}
R_{2}(\theta)A_{bc}(-\theta)B_{c}, \eqno(40)$$
where the amplitudes $R_{1}(\theta)$ and $R_{2}(\theta)$ are shown in
Fig.10.

The boundary Yang-Baxter equation (19) provides three non-trivial equations:
$$R_{1}S_{3}R_{2}S_{1}+R_{2}S_{1}R_{1}S_{1}+(q-3)R_{2}S_{1}R_{2}S_{1}
+(q-3)R_{1}S_{2}R_{2}S_{0}$$
$$+(q-3)R_{2}S_{0}R_{1}S_{0}+(q-3)(q-4)R_{2}S_{0}R_{2}S_{0}$$
$$=(q-2)R_{2}S_{2}R_{1}S_{2}+(q-3)R_{1}S_{0}R_{2}S_{2}+{(q-3)}^{2}
R_{2}S_{0}R_{2}S_{2}$$
$$+R_{1}S_{1}R_{2}S_{3}+R_{2}S_{3}R_{1}S_{3}+(q-3)R_{2}S_{1}R_{2}S_{3};
\eqno(41a)$$
$$R_{1}S_{1}R_{2}S_{2}+(q-3)R_{2}S_{1}R_{2}S_{2}+R_{2}S_{2}R_{1}S_{3}
+(q-3)R_{2}S_{0}R_{2}S_{3}$$
$$+(q-3)R_{2}S_{2}R_{1}S_{2}+R_{2}S_{3}R_{1}S_{2}+(q-3)R_{1}S_{0}R_{2}S_{2}+
(q-3)(q-4)R_{2}S_{0}R_{2}S_{2}$$
$$=R_{1}S_{2}R_{2}S_{1}+(q-3)R_{1}S_{2}R_{2}S_{0}+(q-3)R_{2}S_{0}R_{2}S_{1}+(q-3)^2
R_{2}S_{0}R_{2}S_{0};
\eqno(41b)$$
$$R_{1}S_{3}R_{2}S_{0}+R_{1}S_{2}R_{2}S_{1}+(q-4)R_{1}S_{2}R_{2}S_{0}
+R_{2}S_{1}R_{1}S_{0}+R_{2}S_{0}R_{1}S_{1}$$
$$+(q-4)R_{2}S_{0}R_{1}S_{0}+(q-3)R_{2}S_{1}R_{2}S_{0}+(q-4)R_{2}S_{0}R_{2}S_{1}+
(q-4)^2R_{2}S_{0}R_{2}S_{0}$$
$$=R_{1}S_{1}R_{2}S_{2}+R_{1}S_{0}R_{2}S_{3}+(q-4)R_{1}S_{0}R_{2}S_{2}+
R_{2}S_{3}R_{1}S_{2}+R_{2}S_{2}R_{1}S_{3}$$
$$+(q-3)R_{2}S_{2}R_{1}S_{2}+(q-3)R_{2}S_{1}R_{2}S_{2}+(q-4)R_{2}S_{0}R_{2}S_{3}+
[(q-3)+(q-4)^2]R_{2}S_{0}R_{2}S_{2};
\eqno(41c)$$
where the argument in each term has the form
$R_i(\theta_1)S_j(\theta_1+\theta_2)R_k(\theta_2)
S_l(\theta_2-\theta_1)$.

Equation(41) can be solved for the ratio $R_1(\theta_1)/ R_2(\theta_1)$ when we
take the limit
${\theta_2}\rightarrow {i\pi\over2}$ and noting that both $R_1(\theta_2)$ and
$R_2(\theta_2)$ have a
simple pole in this limit with the same residue
$$Res_{\theta_2 = {i\pi\over2}} R_1(\theta_2) = Res_{\theta_2 = {i\pi\over2}}
R_2(\theta_2) =
{i\over2}g^2 \eqno(42)$$
for some boundary coupling $g$.

The solution to (41) can then be written as
$$R_1(\theta) =
(q-3){\sinh\lambda({i\pi\over3}+2\theta)\over\sinh\lambda(i\pi-2\theta)}P({\lambda
\theta\over i\pi})\eqno(43a)$$
$$R_2(\theta) = {\sin{2\pi\lambda\over3}\over\sin{\pi\lambda\over3}}
{\sinh2\lambda\theta\over\sinh\lambda(i\pi-2\theta)}
{\sinh\lambda({i\pi\over3}+2\theta)\over\sinh\lambda({i\pi\over3}-2\theta)}P({\lambda
\theta\over i\pi}),\eqno(43b)$$
where we use the fact that $R_2(\theta)$ has a simple pole at
$\theta={i\pi\over6}$, which is
absent in $R_1(\theta)$. We do not expect $R_1(\theta)$ and $R_2(\theta)$
to have any other poles
in the physical domain ($0\le\theta\le {i\pi\over2}$). The normalization
factor $P({\lambda\theta\over i\pi})$ is constrained by the unitarity
conditions (20) and (21),
which reduces to
$$P(\theta)P(-\theta) = 1,\eqno(44a)$$
and
$$P({i\pi\over2}-{\theta\over2}) = -\Pi({\lambda\theta\over i\pi})
{\sinh\lambda(i\pi+\theta)\over\sinh\lambda({i\pi\over3}-\theta)}
{\sinh\lambda({4i\pi\over3}+\theta)\over\sinh\lambda({4i\pi\over3}-\theta)}
P({i\pi\over2}+{\theta\over2}),\eqno(44b)$$
respectively. The minimal solutions can be written as
$$P(X) = \prod_{k=1}^{\infty}{\Omega_k(X)\over\Omega_k(-X)},\eqno(45a)$$
where
$$\Omega_k(X) =$$
$${\Gamma[(4k-1)\lambda+2X]\Gamma[(4k-3)\lambda
+2X+1]\Gamma[4k\lambda+{\lambda\over3}-2X]\Gamma[4(k-1)\lambda-{\lambda\over3}-2X+1]}\over
{\Gamma[4k\lambda+2X]\Gamma[4(k-1)\lambda+2X+1]
\Gamma[(4k-3)\lambda+{\lambda\over3}-2X]
\Gamma[(4k-1)\lambda-{\lambda\over3}-2X+1]},\eqno(45b)$$
up to some CDD factors. The sign in (45) can be justified by the boundary
bootstrap equation (24).

In the Ising limit ($\lambda\rightarrow {3\over4}$), the boundary S-matrix
element simplifies
to
$$R_1(\theta) = -\cot({\pi\over4}+{i\theta\over2}),\eqno(46)$$
in agreement with [5]. For the $q=3$ Potts model ($\lambda\rightarrow1$), the
two scattering
amplitudes have the form
$$R_1(\theta) = 0$$
$$R_2(\theta) =
{\sin({\pi\over12}-{i\theta\over2})\sin({\pi\over4}-{i\theta\over2})\over
\sin({\pi\over12}+{i\theta\over2})\sin({\pi\over4}+{i\theta\over2})}.\eqno(47)$$
Finally, a simple computation give us
$$g(\lambda)=-\sqrt{{(3-q)\over \lambda}sin({4\pi\over 3}\lambda)}
exp\lbrace\int_{0}^{\infty}{dt\over t}{e^{-t-4\lambda
t}\over2(1-e^{-t})(1+e^{-2\lambda t})}$$
$$[(1-e^{-\lambda t})(1-e^{3\lambda t-t})+e^{{\lambda
t\over3}-t}(1-e^{-2\lambda t+
{\lambda t\over3}
+t})(1-e^{3\lambda t})]
\rbrace \eqno(48)$$
for the boundary coupling constant.

{\bf Conclusion}

In this study, the q-Potts model boundary S-matrix for free and fixed boundary
conditions
were dervied. It would be interesting to apply the techniques of Thermodynamics
Bethe
Ansatz to study these S-matrices. In particular, one can investigate the
renormalization
group flow between these two boundary conditions[5,13]. This work is in
progress. Finally, we
would like to note that it
may be possible to use these boundary S-matrices to compute the crossing
probabilities
for the percolation problem ($q=1$) in a finite region[12].

{\bf Acknowledgements}

LC was supported by a Eleanor Sophia Wood Travelling Scholarship (Sydney
University)and would like to thank Prof. A.B. Zamolodchikov for many useful
discussions.
\vfill\eject
{\bf References}

1.E.Itzykson, H.Saleur, J.B.Zuber eds., ``Conformal Invariance and Applications

to Statistical Mechanics", World Scientific, 1988.

2.A.A.Belavin, A.M.Polyakov, A.B.Zamolodchikov. Nucl.Phys.B241 (1984), 33.

3.A.B.Zamolodchikov. Advanced Studies in Pure Mathematics 19 (1989), 641.

4.F.Smirnov, ``Formfactors in Completely Integrable Models of QFT'',

Adv. Series in Math. Phys.14, World Scientific, 1992.

5.S.Ghoshal, A.Zamolodchikov. ``Boundary S-Matrix and Boundary State in

Two-Dimensional Integrable Qunatum Field Theory''. Preprint RU-93-20, 1993.

6.R.B.Potts. Proc. Cambridge Phil. Soc. 48 (1952), 106.

7.Vl.S.Dotsenko, V.A.Fateev. Nucl.Phys.B240 (1984) 312.

8.L.Chim, A.Zamolodchikov. Int.J.Mod.Phys. A7, 5317 (1992).

9.J.Cardy. Nucl.Phys.B324 (1989), 581.

10.I.Cherednik. Theor.Math.Phys., 61, 35 (1984) p.977.

11.A.Fring, R.Koberle. ``Factorized Scattering in the
Presence of

Reflecting Boundaries''. Preprint USP-IFQSC/TH/93-06, 1993.

12.J.Cardy. ``Critical Percolation in Finite Geometries''. Preprint
USCBTH-91-56.

13.P.Fendley, H.Saleur. ``Exact Theory of Polymer Adsorption in Analogy

with the Kondo Problem''. Preprint USC-94-006, CM/9403095, 1994.
\vfill\eject
{\bf Figure Captions}
\vskip 0.06in
Fig.1. The scattering processes for the bulk S-matrix elements $S_{0}(\theta),
S_{1}(\theta), S_{2}(\theta)$ and $S_{3}(\theta)$ defined in (9); with
$a\neq b\neq c\neq d$.
\vskip 0.06in
Fig.2. Space-time diagrams associated with the pole at $\theta={{2\pi
i}\over 3}$ in $S_{1}(\theta)$ [Fig.(a)] and the corresponding
cross-channel pole in $S_{2}(\theta)$ [Fig.(b)]. The amplitude
$S_{0}(\theta)$ possess both poles as shown in [Fig.(c)]; the states
$a,b,c,d$ are all different.
\vskip 0.06in
Fig.3. Tree-kink vertex associated with the ``coupling constant'' $f$
\vskip 0.06in
Fig.4. The boundary scattering processes described by the amplitude
$R_{ba}^{c}(\theta)$, with $b\neq a,c$.
\vskip 0.06in
Fig.5. Boundary Yang-Baxter Equation. The variables $g, f, g', f' =
{1,2,...,q}$ satisfies the ``admissibility conditions''
$g\neq c,e$; $f\neq b,g$; $g'\neq a,c$; $f'\neq d,g'$. Boundary conditions
can also place further constraints on these variables.
\vskip 0.06in
Fig.6. Scattering processes described by the boundary unitarity
condition (20); $b\neq a,c$ and $c\neq b$.
\vskip 0.06in
Fig.7. Scattering processes for the cross-unitarity condition (22);
$b\neq a,c$ and $d\neq a,c$.
\vskip 0.06in
Fig.8. Boundary bootstrap equation where states $e$ and $f$ must
satisfy the ``admissibility condition'' and the boundary condition.
\vskip 0.06in
Fig.9. Physical poles of the boundary S-martix with conditions
$a\neq b\neq c$ in [Fig.(a)], and $a\neq b,c$ in [Fig.(b)]. Here
$g_{b}^{c}$ is the boundary coupling constant for the particle $A_{cb}$.
\vskip 0.06in
Fig.10. Boundary S-martix elements for the free boundary condition with
$a, b$, and $c$ all different.
\vfill\eject
\end